\documentclass[12pt,letterpaper]{article}
\usepackage{amsfonts,amsmath,amsxtra}
\usepackage{graphicx}
\usepackage{lscape,epsf}

\newcommand{\bea}{\begin{eqnarray}}
\newcommand{\eea}{\end{eqnarray}}
\newcommand{\be}{\begin{equation}}
\newcommand{\ee}{\end{equation}}
\newcommand{\ba}[1][rcl]{\begin{array}{#1}}
\newcommand{\ea}{\end{array}}
\newcommand{\nn}{\nonumber}

\newcommand{\dlt}{\delta}
\newcommand{\eps}{\epsilon}

\renewcommand{\o}{\omega}

\renewcommand{\t}{\tau}

\newcommand{\Zbb}{\mathbb{Z}}

\newcommand{\Hc}{\mathcal{H}}
\newcommand{\Ic}{\mathcal{I}}

\newcommand{\form}[1]{\mathbf{{#1}}}
\newcommand{\A}{\form{A}}
\newcommand{\B}{\form{B}}
\newcommand{\C}{\form{C}}
\newcommand{\D}{\form{D}}
\newcommand{\E}{\form{E}}
\newcommand{\F}{\form{F}}
\newcommand{\G}{\form{G}}
\newcommand{\J}{\form{J}}
\renewcommand{\H}{\form{H}}
\newcommand{\K}{\form{K}}
\renewcommand{\L}{\form{L}}
\newcommand{\M}{\form{M}}
\newcommand{\N}{\form{N}}
\renewcommand{\P}{\form{P}}
\newcommand{\Q}{\form{Q}}
\newcommand{\R}{\form{R}}
\newcommand{\X}{\form{X}}
\newcommand{\Y}{\form{Y}}

\newcommand{\At}{\form{\tilde{A}}}
\newcommand{\Ct}{\form{\tilde{C}}}

\newcommand{\Ft}{\form{\tilde{F}}}
\newcommand{\Gt}{\form{\tilde{G}}}
\newcommand{\Ht}{\form{\tilde{H}}}
\newcommand{\Nt}{\form{\tilde{N}}}

\newcommand{\thalf}{\frac{\t}{2}}
\newcommand{\half}{\frac{1}{2}}

\newcommand{\h}[1]{h^{#1}}
\newcommand{\sqh}{\sqrt{h}}

\newcommand{\ug}[1]{G^{#1}}
\newcommand{\sg}{\sqrt{-G}}
\newcommand{\wg}{\wedge}
\newcommand{\lpl}{\square_4}
\newcommand{\goverh}{\frac{\ug{55}}{\sqh}}
\newcommand{\mt}{\tilde{m}}

\newcommand{\tr}{\mbox{\,Tr\,}}

\newcommand{\pr}{\partial}
\renewcommand{\d}{\mathrm{d}}
\newcommand{\su}{\mathrm{SU}}

\def\Ah{\hat{A}}

\def\Hh{\hat{H}}
\def\Nh{\hat{N}}

\def\Nn{{\cal{N}}}

\newcommand\Trule{\rule{0pt}{2.6ex}}
\newcommand\Brule{\rule[-1.2ex]{0pt}{0pt}}

\begin{document}

\title{$\Ic$-odd sector of the Klebanov-Strassler theory}
\author{  Anatoly Dymarsky${}^{\,a,\,b}$, Dmitry Melnikov${}^{\,b,\,c}$\footnote{Now at Tel Aviv University, Ramat Aviv 69978, Israel}
\ and Alexander Solovyov${}^{\,d,\,e}$}
\date{}
\maketitle
\thispagestyle{empty}
\begin{center}
\itshape
${}^{a}$ Stanford Institute for Theoretical Physics,
Stanford, CA 94305
\\[1.mm]
${}^{b}$ Institute for Theoretical and Experimental Physics,\\
B.~Cheremushkinskaya, Moscow 117259, Russia
\\[1.mm]
$^{c}$Department of Physics and Astronomy, Rutgers University,
\\ 136 Frelinghuysen Rd, Piscataway, NJ 08854
\\[1.mm]
${}^{d}$ Department of Physics, Princeton University, Princeton, NJ  08544
\\[1.mm]
${}^{e}$  Bogolyubov Institute for Theoretical Physics,
Kiev 03680, Ukraine
\\[1.mm]
\end{center}

\vspace{-13.5cm}
\begin{flushright}
{SU-ITP-08/23}\\
{RUNHETC-2008-17}\\
{PUPT-2280}\\
{ITEP-TH-41/08}\\
{TAUP-2887/08}
\end{flushright}
\vspace{9.0cm}

\begin{abstract}
The Klebanov-Strassler background is invariant under the $\Zbb_2$ symmetry $\Ic$, which acts by exchanging the bi-fundamental fields $A$ and $B$, accompanied by the charge conjugation. We study the background perturbations in the $\Ic$-odd sector and find an exhaustive list of bosonic states invariant under the global $\su(2)\times \su(2)$ symmetry. In addition to the scalars identified in an earlier publication arXiv:~0712.4404 we find 7 families of massive states of spin 1. Together with the spin 0 states they form 3 families of massive vector multiplets and 2 families of massive gravitino multiplets, containing a vector, a pseudovector and fermions of spin~3/2 and~1/2. In the conformal Klebanov-Witten case these $\Ic$-odd particles belong to the $\Nn=1$ superconformal Vector Multiplet I and Gravitino Multiplets II and IV. The operators dual to the $\Ic$-odd singlet sector include those without bi-fundamental fields making an interesting connection with the pure $\Nn=1$ SYM theory. We calculate the mass spectrum of the corresponding glueballs numerically and
discuss possible applications of our results.
\end{abstract}

\newpage

\section{Introduction}
The Klebanov-Strassler supergravity solution, which corresponds to a certain vacuum of the $\su(k(M+1))\times \su(kM)$ gauge theory \cite{KS}, provides an interesting and rich example of the gauge/string duality \cite{Maldacena:1997re,Gubser:1998bc,Witten:1998qj}.
It generalizes the duality between the superconformal $\su(N)\times \su(N)$ gauge theory with bi-fundamentals and string theory on $AdS_5\times T^{1,1}$~\cite{KW}.  Adding extra colors to one of the gauge groups breaks the conformal symmetry~\cite{Gubser:1998fp,KN,KT} and leads to the cascade behavior~\cite{KS,Seiberg duality,Strassler cascade}. The gauge group $\su(k(M+1))\times \su(kM)$ shrinks to $\su(M)$ at the bottom of the cascade and the KS theory reduces to the pure gauge ${\cal N}=1$ SYM \cite{KS}. Unfortunately such a limit requires small $g_s M$, which makes the supergravity approximation invalid. Nevertheless this connection between the KS solution and the pure super-Yang-Mills theory strongly motivates the studies of the bi-fundamental free sector of the $\su(k(M+1))\times \su(kM)$ theory that survives at the bottom of the cascade.

 The KS solution is invariant under the $\Zbb_2$ symmetry  $\Ic$, which acts by exchanging the two two-spheres  of the deformed conifold accompanied by the inversion of sign of the 3-form flux. On the field theory side this symmetry exchanges and conjugates the bi-fundamental fields $A$ and $B$. Thus the KS solution corresponds to one particular  $\Ic$-invariant vacuum  $|A|^2=|B|^2$. The latter spontaneously breaks $U(1)_{\rm Baryon}$ symmetry $A\rightarrow Ae^{i a}$, $B \rightarrow B e^{-i a }$. The corresponding massless Goldstone pseudoscalar $a$ combines with the scalar $U \sim |A|^2-|B|^2$ into a $\Ic$-odd scalar supermultiplet \cite{GHK}. While $a$ corresponds to the longitudinal part of the $U(1)_{\rm Baryon}$ current $J_\mu =\partial_\mu a$, the fluctuation of $U$ changes the expectation values of the baryon operators ${\cal A}$, ${\cal B}$ and moves the theory along the baryonic branch of the moduli space \cite{GHK,Butti,DKS,BDK}.

The massless $\Ic$-odd supermultiplet $(U,a)$ was first studied in \cite{GHK}. Later this analysis was generalized to the massive excitations in \cite{BDKS}. In particular it was shown  that the massive excitations of $U$ mix with another $\Ic$-odd scalar $\chi$, which comes from the NS-NS sector. It was also suggested there that the massive states of the pseudoscalar $a$ are  eaten by a gauge vector and form a massive vector state similarly to the Goldstone boson associated with the chiral symmetry breaking in the Sakai-Sugimoto model \cite{Sakai}. The massive vector is dual to the baryonic current $J_\mu$, and it generalizes the  massless Betti vector of the conformal theory. The massive modes of  $J_\mu$ together with the massive modes of $U$ combine into a tower of massive vector supermultiplets. To accommodate the mixing between $U$ and $\chi$ the Betti vector must mix with another massive vector also from the NS-NS sector. The resulting spectrum of the coupled vector system must coincide with that one of the scalar system $(U,\chi)$ found in \cite{BDKS}.

The symmetry between the NS-NS and RR sectors in the conformal case suggests that in addition to the scalar $\chi$ there should be another scalar $\tilde{\chi}$ from the R-R sector.
The later was found in \cite{BDKS} to decouple from all other excitations in the KS case. It was also conjectured there that $\tilde{\chi}$ is a superpartner of yet another massive vector, which also comes from the R-R sector.

This picture with three massive vectors dual to $U$, $\chi$ and $\tilde{\chi}$  was supported in \cite{BDKS} by a calculation in the simplified Klebanov-Tseytlin background. Although the particles in question form three complete supermultiplets (we have in mind only  bosonic states here) there must be more bosonic states in the $\Ic$-odd sector. Indeed the analysis of the conformal Klebanov-Witten case reveals that the three scalars and three vectors do not complete a representation of the superconformal symmetry.

In this paper we find all other $\Ic$-odd bosonic states in the full KS background. Starting with the most general $\Ic$-odd ansatz in the $\su(2)\times \su(2)$ singlet sector we find that the three massive vectors, predicted in \cite{BDKS}, mix with other four  massive spin 1 states. This leads to the appearance of two new massive supermultiplets in the spectrum, each containing a vector, a pseudovector and two fermions of spin $1/2$ and $3/2$. The three spin 0 and seven spin 1 particles completely span the set of the bosonic components of the shortened Vector Multiplet I and Gravitino Multiplets II and IV~\cite{Ce1,Ce2} of the superconformal KW theory. In this way the spectrum of the $\Ic$-odd bosonic $\su(2) \times \su(2)$-invariant supergravity excitations over the KS background is fully covered.

The comparison with the conformal case suggests that the lightest vector multiplet
(and the corresponding tower) is created by the operators from the composite superfield $\tr Ae^V\bar{A}e^{-V} - \tr Be^V\bar{B}e^{-V}$, while the rest of the massive states are created by the pure super-Yang-Mills operator $\tr e^V\bar{W}_{\dot{\alpha}}e^{-V}W^2$. The latter do not contain bi-fundamental fields. Therefore the spectrum of the corresponding glueballs, which we study numerically, might shed a light on the dynamics of the glueball states in the pure ${\cal N}=1$ SYM theory. We further speculate on this point in section 6.

This paper is organized as follows. In the next section we write down the general $\Ic$-odd singlet ansatz and discuss its relation to the conformal case. Then we find and solve the corresponding equations of motion in sections 3 and 4. The fifth section is devoted to the analysis of the results obtained in the preceding sections. Numerical calculation of the spectra for the new multiplets is followed by the discussion of their quantum numbers and peculiarities in the procedure of calculating scaling dimensions. In the end of section~5 we also discuss the field theory operators dual to the $\Ic$-odd sector. Section 6 concludes the paper with a discussion of the results.

\section{$\Ic$-odd excitations over the KS background}
\subsection{General ansatz}
\label{sec:general_ansatz}
We consider the $\Ic$-odd supergravity excitations over the KS background which are singlets w.r.t.\ the action of $\su(2)\times \su(2)$ $R$-symmetry group. The  $\Ic$-symmetry of the KS solution acts on the conifold geometry by interchanging the two spheres $(\theta_1,\phi_1)$ and $(\theta_2,\phi_2)$, simultaneously changing the sign of $F_3$ and $H_3$. Hence we are looking for the  perturbations of $B_2$ and $C_2$ invariant under the exchange of the two-spheres and the perturbations of metric and $C_4$ which are odd under $(\theta_1,\phi_1) \leftrightarrow (\theta_2,\phi_2)$.

The list of  $\su(2)\times \su(2)$-invariant forms on the conifold include the one-form $\d\t$ along the radius and invariant forms on the ``base'' of the deformed conifold $T^{1,1}$. There is a unique invariant one-form $g^5$, which is $\Ic$-even. It satisfies
\bea
\label{lap1}
\star\, \d \star\, \d g^5= 8 g^5.
\eea
Below $\star$ will denote the Hodge operation on $T^{1,1}$, while $\ast$ and $\ast_4$ will refer to the same operation in the ten-dimensional or four-dimensional spaces respectively.

There are three $\Ic$-odd $\su(2)\times \su(2)$ invariant two-forms:
$g^1 \wg g^2$, $g^3 \wg g^4$ and ${g^1}\wg~{g^3}~+~g^2\wg g^4$ (see~\cite{Tsimpis} for definitions). In addition there are two $\Ic$-even two-forms $\d g_5=-(g^1\wedge g^4+g^3\wedge g^2)$ and $\d \t\wedge g^5$, which are not independent. Any fluctuation including $\d g_5$ can be transformed into the fluctuation with $g_5$ or $\d\t\wedge g^5$ with the help of a suitable gauge transformation.

The invariant two-forms mentioned above can be combined into two eigenvectors of the Laplace-Beltrami operator $\star\, \d$ on $T^{1,1}$ as follows:
\bea
\omega_2 &=& g^1\wedge g^2+g^3\wedge g^4\ , \qquad \d \star\, \omega_2=0\ ,\qquad \d \omega_2=0\ ,\\
Y_2 &=&(g^1\wedge g^2-g^3\wedge g^4)+i(g^1\wedge g^3+g^2\wedge g^4)\ ,\qquad
\d \star\, Y_2=0,\quad \star\, \d Y_2= 3i\, Y_2.\quad
\label{lap2}
\eea
There are also three and four forms on $T^{1,1}$ invariant under $\su(2)\times \su(2)$, but they all can be obtained from the forms above using the exterior differentiation and the Hodge transformation.
The only $\Ic$-odd $\su(2)\times \su(2)$-invariant metric fluctuation is $g^1\cdot g^2 + g^3\cdot g^4$.

The Hodge duality in Minkowski space allows one to relate the $p$- and $(4-p)$-forms to each other. That is why the general ansatz can be written in terms of  zero, one and two-forms in Minkowski space. It is also known that any form has a Hodge decomposition into the sum of an exact, co-exact and harmonic parts. The field theory in the $\Zbb_2$-symmetric vacuum dual to the KS background does not have any spontaneously broken symmetries besides $U(1)_{\rm Baryon}$. Therefore we do not expect any $\su(2)\times \su(2)$ singlet massless particles in addition to those associated with the baryonic branch of the moduli space. The latter were studied in \cite{GHK,BDKS,Argurio}. That is why we are looking only for massive excitations, i.e. all four-dimensional forms $P_k$ in our ansatz satisfy
\bea
\lpl P_k = m^2 P_k
\eea
with some non-zero $m^2$. It means that the harmonic part is absent from the decomposition (which is not generally the case for the four-dimensional massless modes). Therefore, any two-form $P_2$ can be written using the two vectors (one-forms) $\M$ and $\N$:\footnote{We use the boldface notation for the spin 1 excitations throughout the paper.}
\bea
P_2 &=& \d_4 \M + \ast_4 \d_4 \N \,.
\eea
Similarly, any vector $\N$ can be represented as a sum of an exact and a co-closed parts:
\bea
\N &=& \d_4 \chi + \Nt \,,
\eea
where
\bea
\d_4 \ast_4 \Nt &=& 0 \,.
\eea

This consideration shows that all the $\Ic$-odd excitations over the KS background reduce to some vector and scalar ansatz. At this point we do not make a distinction between the particles with different behavior with respect to parity;  i.e. vectors and axial vectors, scalars and axial scalars. For the sake of simplicity we call all states of spin 1 ``vectors'' and all states of spin 0 ``scalars''. The quantum numbers of the physical states, including parity, are given in figure~\ref{fig spectrum} in section \ref{numerics sec}.

The most general scalar ansatz was considered in~\cite{BDKS}.
Namely, there are the following two decoupled systems of excitations,
\be\label{ansatzH}
\ba
\delta B_2 &=& \chi(x,\tau)\, \d g^5 + \pr_\mu \sigma(x,\tau)\, \d x^\mu \wg g^5 \,,
\\
\delta G_{13} &=& \delta G_{24} = U (x,\tau) \ ,
\ea
\ee
and
\bea
\label{ansatzF}
\delta C_2 &=& \tilde{\chi}(x,\tau)\, \d g^5 + \pr_\mu \tilde{\sigma}(x,\tau)\, \d x^\mu \wg g^5 \,.
\eea
As it  was mentioned above, the terms proportional to $\d\t\wedge g^5$ are absent because they can be transformed into (\ref{ansatzH}) and (\ref{ansatzF}) with  help of a gauge transformation.

One could seemingly add the $\Ic$-odd scalar excitations of $F_5$,
\bea
\dlt F_5 &=& (1+\ast) \bigl[ \d\t \wg (\d_4 a \wg g^1 \wg g^2 + \d_4 b \wg g^3 \wg g^4) \wg g^5 \bigr] \,;
\eea
or
\bea
\dlt F_5 &=& (1+\ast) \bigl[ \d_4 c \wg \d\t (g^1 \wg g^3 + g^2 \wg g^4) \wg g^5 \bigr] \,.
\eea
However, equations of motion would require the functions $a$, $b$ and $c$ to vanish identically.\footnote{Note that this is not the case for the massless particles~\cite{GHK}.} After some redefinition of the variables the equations of motion become
\bea
\label{ansHeq1}
{z}''-\frac{2}{\sinh^2 \tau}\,{z}+
\tilde m^2\, \frac{ I(\tau)}{ K^2(\tau)}\,
{z} &=& 2^{2/3}\mt\, K(\tau)\, {w}\ ,
\\
\label{ansHeq2}
{w}''-\frac{\cosh^2\tau+1}{\sinh^2 \tau}\,
{w}+ \tilde m^2 \,\frac{ I(\tau)}{ K^2(\tau)}\, {w} &=&
2^{2/3}\mt\, K(\tau)\, {z}
\eea
for (\ref{ansatzH}) and
\bea
\label{ansFeq}
\tilde{w}''-\frac{\cosh^2\tau+1}{\sinh^2 \tau}\,
\tilde{w}+\mt^2 \frac{I(\tau)}{K(\tau)^2}\, \tilde{w}=0
\eea
for (\ref{ansatzF}) respectively \cite{BDKS}. The definitions of the background functions $K(\t)$, $I(\t)$ in the above expressions can be found in the appendix~\ref{app:KS}, while the four-dimensional mass normalization $\mt$ is defined by~(\ref{mt}).

The most general $\su(2)\times \su(2)$ singlet $\Ic$-odd vector excitation of the 3-form potentials is as follows:
\bea
\label{gen C2v}
\C^{(1)}\wedge \d\tau + \C^{(2)}\wedge g^5 + \ast_4\d_4\C^{(3)}.
\eea
For the 5-form the most general vector perturbation is
\bea
\nn
&&(1+*)\,\left[\F^{(1)}\wedge \d\tau\wedge g^5\wedge g^1\wedge g^2 + \F^{(2)}\wedge \d\tau\wedge g^5\wedge g^3\wedge g^4 \right. +
\\ \label{gen F5v}
&+& \F^{(3)}\wedge \d\tau\wedge g^5\wedge (g^1\wedge g^3 + g^2\wedge g^4) + (\d_4\F^{(4)}+*_4\d_4\F^{(5)})\wedge  g^5\wedge g^1\wedge g^2 +
\\ \nn
&+& \left.  (\d_4\F^{(6)}+*_4\d_4\F^{(7)})\wedge  g^5\wedge g^3\wedge g^4 + (\d_4\F^{(8)}+*_4\d_4\F^{(9)})\wedge  g^5\wedge (g^1\wedge g^3+g^2\wedge g^4)\right]\ .
\eea
Not all fifteen (3+3+9) real vectors above are independent. This ansatz has only seven independent vector degrees of freedom. We show this in the next section by considering the conformal KW case.

\subsection{Supermultiplet structure in the conformal case}
\label{KW modes sec}
We start our analysis with the scalar $U$ of \cite{GHK,BDKS} dual to the operator ${\rm Tr}\left(|A|^2-|B|^2\right)$ of dimension~2~\cite{DKS}. In the conformal case this operator is responsible for the resolution of the conifold. The corresponding state belongs to the Betti multiplet~\cite{KWII}. The latter also contains a 5d-massless gauge vector of dimension 3 dual to the baryonic current. Its presence on the gravity side is guaranteed by the nontrivial harmonic three-form $w_3~=~\star~w_2$ on $T^{1,1}$. The Betti multiplet is a ``massless'' Vector Multiplet~I according to the classification of the superconformal multiplets given in~\cite{Ce1,Ce2}. It is a short version of the Vector Multiplet I, which contains just two bosonic states of dimensions 2 and 3.

\begin{table}[htb]

\begin{center}
\caption{\small Shortened Gravitino Multiplets II, IV (left) and Vector Multiplet I (right)~\cite{Ce1,Ce2}. Field notations are inherited from~\cite{KRN}.}
\label{GMIV}
\begin{tabular}[c]{@{\hspace{0.0cm}}c@{\hspace{0.2cm}}c}\small
 \begin{tabular}[c]{||c|c|c|c|c|| }\hline \Trule\Brule
  Field & reps & $\Delta$ & ${\cal{R}}$ & Mode
\\ \hline \Trule\Brule
$a_\mu$ & $(1/2,1/2)$ & 5 & 0 & $\C^{(2)}$, ($\chi$, $\tilde{\chi}$)
\\ \hline
$b_{\mu\nu}^\pm$ & $(1,0)$, $(0,1)$ & 5 & $\mp 2$ & $\F^{(1)}-\F^{(2)}$, $\F^{(3)}$
\\ \hline
$a_{\mu\nu}$ & $(1,0)$, $(0,1)$ & 6 & 0 & $\C^{(3)}$
\\ \hline
\end{tabular}
& \small
 \begin{tabular}[c]{||c|c|c|c|c||}\hline \Trule\Brule
 Field & reps & $\Delta$ & ${\cal{R}}$ & Mode
\\ \hline \Trule\Brule
 $\phi_\mu$ & $(1/2,1/2)$ & $3$ & 0 & $\F^{(1)}+\F^{(2)}$
\\ \hline
 $\phi$ & $(0,0)$ & $2$ & 0 & U
\\ \hline
\multicolumn{5}{c}{}
\end{tabular}
\end{tabular}
\end{center}
\end{table}

In the table~\ref{GMIV} we match the components of the five-dimensional superconformal multiplets of~\cite{Ce1} to the four-dimensional fluctuations  considered in the previous section. The identification of $U$ as $\phi$ from the table~\ref{GMIV} is straightforward. The Betti vector $\phi_\mu$,
\bea
\delta C_4=\phi_\mu \wedge \omega_3 \,,
\eea
is contained in (\ref{gen F5v}). The combination $\F^{(1)}+\F^{(2)}$ is identified with the derivative of $\phi_\mu$ with respect to $\tau$ and the remaining functions $\F^{(3)},..,\F^{(9)}$ are dependent on $\F^{(1)}+\F^{(2)}$.

The scalars $\chi,\tilde{\chi}$
have dimension $5=2+\sqrt{1+{\bf 8}}$ as it follows from (\ref{lap1}). The same result
follows from the large $\tau$ behavior of (\ref{ansHeq2}) and (\ref{ansFeq}). Consequently $\chi,\tilde{\chi}$ are the  longitudinal modes of the five-dimensional vectors $a_\mu$ from the Gravitino Multiplets of type~II and~IV.
It is convenient to consider these multiplets together combining the modes into the complex combinations like
\bea
\label{c2b2}
\delta B_2+i\,\delta C_2=a_\mu\wedge g^5\ .
\eea

Similarly the complex vector $\C^{(2)}$ from (\ref{gen C2v}) corresponds to the vector part of~$a_\mu$. It has dimension 5 in the KW case as well. The complex vectors $\C^{(1)},\C^{(3)}$ correspond to the antisymmetric tensor $a_{\mu\nu}$ and have dimension $6$. Only one of them is independent on-shell.

Although the fluctuations of the RR four-form $C_4$ are real they can be parameterized
with help of complex $b_{\mu\nu}$,
\bea
\label{dC4}
\delta C_4=b_{\mu\nu}\wedge Y_2 +c.c.\ .
\eea
By comparing (\ref{dC4}) to (\ref{gen F5v}) we identify the real components of $b_{\mu\nu}$ with $\F^{(1)}-\F^{(2)}$ and $\F^{(3)}$. All other vectors $\F^{(4)},..,\F^{(9)}$ are not independent on-shell.
These fluctuations have dimension $5=2+|{\bf 3}|$ due to (\ref{lap2}) and also belong to the Gravitino Multiplets II and IV.

In the $\su(2)\times \su(2)$ invariant sector only shortened version of the Gravitino Multiplets II and IV appear. Thus we do not expect any other massive bosonic states in the $\Ic$-odd sector. This agrees with our study of the ansatz (\ref{gen C2v}), (\ref{gen F5v}) in the following section.

There are two ways one can look at the system given by the vector ansatz (\ref{gen C2v}), (\ref{gen F5v}) and the scalar ansatz~(\ref{ansatzH}), (\ref{ansatzF}). First, one can classify the states according to the complex representation of the superconformal symmetry. Second, one can look for the states of definite parity. The second approach is more straightforward. In particular, as it is demonstrated in~\cite{BDKS} the definite parity R-R and NS-NS sectors decouple from each other, although they are a mixture of  states from the superconformal Gravitino Multiplets. Therefore instead of dealing with the Gravitino Multiplets~II and~IV independently we will refer to the combination of the Gravitino Multiplets II and IV just as to the ``Gravitino Multiplets'' and specify the parity where appropriate. That is why we combined the Gravitino Multiplets II and IV together in the table~\ref{GMIV}.

More precisely, the study of the KS case in \cite{BDKS} implies the following. The scalar $U$ from the Vector Multiplet I mixes with the scalar $\chi$ from NS-NS sector of the Gravitino Multiplets, while the pseudoscalar $\tilde{\chi}$ from the R-R sector decouples.
At the same time the calculation done there in the large $\t$ approximation shows that the  Betti pseudovector mixes with the pseudovector part of $a_\mu$ from the R-R sector, though both decouple from the vector part of $a_\mu$ from the NS-NS sector. This suggests that the vector excitations from the Gravitino Multiplets and the Vector Multiplet I split into the following two non-interacting systems. One includes the spin 1 states of positive parity from $a_\mu$, $a_{\mu\nu}$ (NS-NS sector) and  one of the $b_{\mu\nu}$ modes. Another consists of the spin 1 states of negative parity and includes the vectors from  $a_\mu$, $a_{\mu\nu}$ (R-R sector) and another $b_{\mu\nu}$ mode together with the Betti pseudovector.

\section{Triplet of vectors from the Gravitino Multiplets}
\label{3-system sec}
This section analyzes the vector fluctuations  from the Gravitino Multiplets, more precisely a combination of the Gravitino Multiplets II and IV with negative parity. The system of the linearized equations in this subsector reduces to three coupled equations, which can be disentangled. Here we present only the results of our analysis. The reader can find more details of the calculations in the appendix~\ref{app;derivation}.

\subsection{Derivation of the equations}
We start with writing down a general ansatz for the spin 1 excitations in the ``NS-NS sector'' of the Gravitino Multiplets  and show that they decouple from the other vectors.
The deformations of the three and five-forms are:
\bea
\label{ansVH1}
\dlt B_2 &=& \ast_4 \d_4 \H + \A \wg g^5 \,,
\\
\label{ansVF1}
\dlt C_2 &=& \E \wg \d\t \,,
\eea
\bea
\label{deltaF5_1}
\nn \dlt F_5 &=& (1+\ast)\, \bigl[ \d_4 \K \wg \d\t \wg g^1\wg g^2 + \d_4 \L \wg \d\t \wg g^3\wg g^4
\\
&& + \d_4 \M \wg (g^1\wg g^3 + g^2 \wg g^4) \wg g^5 + \N \wg \d\t \wg (g^1 \wg g^3 + g^2 \wg g^4) \wg g^5 \bigr]\,.
\eea
As it was discussed in section 2, vector $\A$ corresponds to $a_\mu$,
vectors $\E$ and $\H$ to $a_{\mu\nu}$ and $\K$, $\L$, $\M$, $\N$ to $b_{\mu\nu}$ of the conformal case.
The equations of motion presented in the appendix B show that $\E$, $\K$, $\L$ and $\M$
depend on the $\A$, $\H$, $\N$ algebraically. The latter describe the physical degrees of freedom.
After redefining $\N$ and $\A$:
\bea
\label{Nt}
\goverh\, \N &=& \lpl \Nt \ ,\\
\label{At}
  K^2 \sinh\t\, \A &=& \At \,;
\eea
the resulting equations take the form:
\begin{multline}
\label{Nteq}
\Nt'' - \left( \frac{\cosh^2\t +1}{\sinh^2\t}  + \frac{4\cdot 2^{1/3}(F')^2}{I K^2 \sinh^2\t} \right) \, \Nt + \mt^2 \frac{I}{K^2}\, \Nt + \\
+ F'\H' - \frac{2^{1/3}F' \ell}{I K^2 \sinh^2\t}\, \H + \frac{F'}{K^2\sinh\t}\, \At = 0 \,,
\end{multline}
\bea
\label{Ateq}
\At'' - \frac{\cosh^2\t+1}{\sinh^2\t}\, \At + \mt^2 \frac{I}{K^2} \At + \mt^2 \frac{4 \cdot 2^{1/3} F'}{K^2 \sinh\t}\, \Nt = 0\,,
\eea
\begin{multline}
\label{Heq}
\H'' + \left( 2\frac{\bigl(K\sinh\t \bigr)'}{K\sinh\t} + \frac{I'}{I} \right)\H' -\left( \frac{2^{1/3} \ell'}{I K^2\sinh^2\t} + \frac{2^{2/3} \ell^2}{I^2 K^4 \sinh^4\t} \right) \H  + \mt^2 \frac{I}{K^2}\, \H -
\\
- \frac{4\cdot 2^{1/3}}{I K^2 \sinh^2\t}\, \bigl( F'\Nt \bigr)' - \frac{4\cdot 2^{2/3} F' \ell}{I^2 K^4 \sinh^4\t}\, \Nt = 0 \,.
\end{multline}

Our goal here is to diagonalize the above system. In particular, we expect to identify the massive vector superpartner of the scalar (\ref{ansatzF}).

\subsection{Analysis of the equations}
\label{3-system ansec}
Although the equations (\ref{Nteq})-(\ref{Heq}) look bulky it is quite easy to split them to three independent equations. First we notice that the constraint $\Nt=0$
implies
\bea
\label{HAt}
\H=\frac{(\sinh\tau \At)'}{\tilde{m}^2 I \sinh^2 \tau}\ ,
\eea
and it reduces the system (\ref{Nteq})-(\ref{Heq}) to one equation
\bea
\label{Aeq_same}
\At^{\prime\prime} - \frac{\cosh^2\tau + 1}{\sinh^2\tau}\,\At + \mt^2 \frac{I}{K^2}\,\At = 0\ .
\eea
This equation coincides with the one for the scalar $\tilde{\chi}$ (\ref{ansFeq}). Hence the vector mode above and the scalar (\ref{ansFeq}) form a massive  vector $j=1/2$  multiplet\footnote{We use spin $j$ to characterize the  massive supermultiplets $(j-1/2)\oplus j \oplus j \oplus (j+1/2)$.} as was predicted in \cite{BDKS}.
It is interesting to notice that $\Nt=0$ does not imply $\delta F_5=0$ as it would in the conformal case. Rather
$F_5 = (1+\ast)\, \d_4 \H \wg H_3$  with $\H$ being related to $\At$ by (\ref{HAt}).

To find the two remaining modes we impose a constraint
\bea
\label{1st order3}
\Ht=- \frac{K(\sinh \tau  \At)'}{ \mt^2 \sqrt{I}\sinh\t}\ ,
\eea
where $\Ht=\sqrt{I} K\sinh\t\, \H$.  This constraint guarantees that the two remaining modes are orthogonal to the vector mode from above. More details on the disentanglement procedure can be found in the appendix~\ref{app;derivation}. Eliminating $\Ht$ from the above equations one obtains
\bea
\label{Aheq1}
\At^{\prime\prime} - \frac{\cosh^2\tau+1}{\sinh^2\tau}\,\At + \mt^2\,\frac{I}{K^2}\,\At -\frac{2\mt^2 I^\prime}{K^3\sinh\tau}\,\Nt =0 \,,
\\
\label{Nheq1}
\Nt^{\prime\prime} - \frac{\cosh^2\tau+1}{\sinh^2\tau}\, \Nt + \mt^2\,\frac{I}{K^2}\,\Nt - \frac{2^{-1/3}I^\prime}{K^3\sinh\tau}\,\At  =0 \,.
\eea
After a trivial rescaling and change of variables  $\X_\pm=\At\pm 2^{2/3}\tilde{m}\Nt$ these two equations decouple,
\bea
\label{Xeqs}
\X_\pm^{\prime\prime} - \frac{\cosh^2\tau+1}{\sinh^2\tau}\, \X_\pm + \mt^2\,\frac{I}{K^2}\, \X_\pm \mp \frac{2^{5/3}\mt F^\prime}{K^2\sinh\tau}\, \X_\pm  =0 \,.
\eea
In the next section  we are going to show that these particles are members of the two $j=1$ gravitino multiplets and find their vector superpartners.

\section{Betti vector and axial vector triplet}
\label{4-system sec}

In this section we consider the vector excitations in the parity even ``R-R sector'' of the combination of the Gravitino Multiplets and
the axial Betti vector from Vector Multiplet I. We expect this system of four vectors to contain the superpartners of the scalar excitations (\ref{ansatzH}) and the two vectors $\X^\pm$ from section 3.

\subsection{Derivation of the equations}
We consider the following deformations of the 3-form potentials
\bea
\label{ansVH2}
\dlt B_2 &=& \J \wg \d\t \,,
\\
\label{ansVF2}
\dlt C_2 &=& \C \wg g^5 + \ast_4 \d_4 \D \,,
\eea
and the 5-form
\begin{multline}
\label{deltaF5_2}
\dlt F_5 = (1+\ast) \bigl[ \F\wg \d\t \wg g^1 \wg g^2 \wg g^5 + \G \wg \d\t \wg g^3 \wg g^4 \wg g^5
\\
 + \d_4\P \wg g^1 \wg g^2 \wg g^5 + \d_4\Q \wg g^3 \wg g^4 \wg g^5 + d_4 \R \wg \d\t \wg (g^1\wg g^3 + g^2 \wg g^4) \bigr] \,.
\end{multline}
Clearly in the conformal limit $\C$ corresponds to $a_\mu$, while $\J$ and $\D$ to $a_{\mu\nu}$.  The fluctuations of $F_5$ correspond to both $b_{\mu\nu}$ and $\phi_\mu$.

Choosing $\C,\D,\F$,and $\G$ as independent variables we
end up with the following system of four coupled equations:
\bea
\label{B+eq}
\B_+'' -  \frac{2}{\sinh^2\tau}\, \B_+ + \mt^2 \frac{I}{K^2}\, \B_+ +  K^3\sinh\tau (\D' -  \J) - K\Ct  &=& 0\,,
\\
\label{B-eq}
\B_-'' - \frac{\cosh^2\tau+1}{\sinh^2\tau}\, \B_- + \mt^2 \frac{I}{K^2}\, \B_-  + 2^{-1/3}\frac{I^\prime}{K}\,(\D' - \J)  + \frac{2^{-1/3}I^\prime}{K^3 \sinh\tau}\, \Ct &=& 0\,,
\\
\label{CeqIKS2}
\Ct'' - \frac{\cosh^2\t+1}{\sinh^2\t}\, \Ct + \mt^2 \frac{I}{K^2}\, \Ct -2^{1/3}\mt^2K\B_+ +\mt^2\,\frac{I^\prime}{K^3 \sinh\tau}\,\B_- &=& 0,
\\
\label{DeqIKS2}
\D'' + \Bigl( \log(IK^2\sinh^2\t) \Bigr)' \D' + \mt^2 \frac{I}{K^2}\, \D +\frac{(I^\prime K^2\sinh^2\tau)^\prime}{IK^2\sinh^2\tau}\,\D    \ + \qquad\quad && \nn \\
+ \frac{I^\prime}{I}\,\J - \frac{1}{IK^2\sinh^2\tau}\,  \left( 2^{1/3}K^3\sinh\tau\,\B_+ + \frac{I^\prime}{K}\,\B_- \right)' &=& 0 \,;
\eea
where
\bea
\label{Jeq2}
\J&=& -\frac{I^\prime}{I}\,\D + \frac{2^{1/3}K}{I\sinh\tau}\,\B_+  + \frac{I^\prime}{IK^3\sinh^2\tau}\B_- \,.
\eea
Here we introduced new variables as follows:
\bea
\label{Ft}
\goverh\,  \coth^2\thalf\, \F &=& \coth\thalf\, \lpl\Ft \,,
\\
\label{Gt}
\goverh\, \tanh^2\thalf\, \G &=& \tanh\thalf\, \lpl\Gt \,,\\
\B_{\pm} &=& \Ft\pm\Gt\ ,\\
\tilde{\C}&=& K^2 \sinh\tau \C\ .
\eea

\subsection{Analysis of the equations}
The system of the equations (\ref{B+eq})-(\ref{DeqIKS2}) can be further reduced.
The hint is to consider a conformal limit when the Betti vector decouples from the
Gravitino Multiplet states. The former is associated with $\F=\G$ while the
perturbation $b_{\mu\nu}$ from the Gravitino Multiplet corresponds to $\F=-\,\G$.
We put
\bea
\label{constraint2}
\B_- &=& 0 \,,
\eea
to ``turn of'' the excitation of $b_{\mu\nu}$ in the system
(\ref{B+eq})-(\ref{DeqIKS2}). This
implies for $\D$
\bea
\label{Ctprimeeq}
\D=\frac{(\sinh \tau\, \Ct)'}{ \mt^2 I\sinh^2\tau}\ .
\eea
The remaining equations form a self-consistent subsystem of two equations:
\bea
\label{B+eqIKS3}
\B_+'' -  \frac{2}{\sinh^2\tau}\,\B_+ + \mt^2 \frac{I}{K^2}\, \B_+  &=& 2K\Ct  \,,
\\
\label{CeqIKS3}
\Ct'' - \frac{\cosh^2\t+1}{\sinh^2\t}\, \Ct + \mt^2 \frac{I}{K^2}\, \Ct &=&2^{1/3}\mt^2K\B_+  \,.
\eea
After a trivial rescaling of variables it reproduces the scalar equations (\ref{ansHeq1}) and (\ref{ansHeq2}). Thus these modes represent the mixing of the Betti vector with the vector part of $a_\mu$. They are the vector superpartners of the scalar excitations $z$ and $w$ discovered in \cite{BDKS}.

To extract the remaining degrees of freedom we ``turn off'' the Betti vector by choosing
\bea
\B_+ &=& 0 \ .
\eea
Using this equation one can eliminate $\D$ from the remaining equations
\bea
\label{Ctprimeeq2}
\D=-\frac{(\sinh \tau\, \Ct)'}{\mt^2 I\sinh^2\tau}\ .
\eea
The remaining self-consistent subsystem of the two equations for $\B_-$ and $\Ct$ is
\bea
\label{B-eqIKS4}
\B_-'' - \frac{\cosh^2\tau+1}{\sinh^2\tau}\, \B_- + \mt^2 \frac{I}{K^2}\, \B_-  &=& - \frac{2^{2/3}I^\prime}{K^3 \sinh\tau}\, \Ct \,,
\\
\label{CteqIKS4}
\Ct'' - \frac{\cosh^2\t+1}{\sinh^2\t}\, \Ct + \mt^2 \frac{I}{K^2}\, \Ct &=& -\mt^2\,\frac{I^\prime}{K^3 \sinh\tau}\,\B_- \,.
\eea
After a trivial rescaling and change of variables $\Y_\pm = 2^{-1/3}\tilde{m}\B_- \mp \Ct$ the equations become
\bea
\label{Xeqs2a}
\Y_\pm'' - \frac{\cosh^2\tau+1}{\sinh^2\tau}\, \Y_\pm + \mt^2 \frac{I}{K^2}\, \Y_\pm  \mp \frac{2^{5/3} \mt F^\prime}{K^2 \sinh\tau}\, \Y_\pm &=& 0 \,.
\eea
These equations exactly coincide with the system (\ref{Xeqs}), which suggests that we have found the members of the same supermultiplets.
Namely, we have the two $j=1$ supermultiplets each containing a vector $\X$, an axial vector $\Y$ and two fermions of spin $1/2$ and $3/2$.

The spectra of the vector supermultiplets, which include scalars, were found in \cite{BDKS}. We devote section~\ref{numerics sec} to the numerical study of the spectra of the $j=1$ multiplets (\ref{Xeqs}).

\section{Analysis of the results}

\subsection{Numerical calculation of spectra}
\label{numerics sec}
In this work we only need to compute the spectra of the decoupled differential equations (\ref{Xeqs}) or (\ref{Xeqs2a}), which describe four vector excitations $\X_{\pm},\Y_{\pm}$ found above. The spectra of the other three vector excitations are the same as of their scalar superpartners (\ref{ansHeq1}), (\ref{ansHeq2}) and (\ref{ansFeq}). They were already  computed in \cite{BDKS}.

We follow the conventions of the work \cite{BDKS}, which uses the following definition of the warp-factor:
\be
h(\tau) = 4 \cdot 2^{2/3} \eps^{-8/3} I(\t)\,,
\ee
where
\be
I(\tau) \equiv
\int_\tau^\infty \d x \,\frac{x\coth x-1}{\sinh^2 x}\, (\sinh (2x) - 2x)^{1/3} \,.
\ee
The eigenvalues are computed in units of $\mt^2$, defined in (\ref{mt}). The shooting method for equations (\ref{Xeqs}) gives the two spectra, listed in the table~\ref{tab:eigenvals1}. In the units, used by Berg~et.al.~\cite{BHM1} the lowest states have masses
$$
\mt^2_{-} = 1.78;
\qquad
\mt^2_{+} = 2.83.
$$

\begin{table}[tb]
\begin{center}
\caption{\small
Lowest values of $\mt^2$ and quadratic fit for the $j=1$ multiplets described by (\ref{Xeqs}).}
\label{tab:eigenvals1}
\begin{tabular}{cc}
{\small Spectrum of $\X_-$ } & {\small Spectrum of $\X_+$}
\\
$\mt^2_{-} = 0.633 + 1.02\, n + 0.287\, n^2$ &  $\mt^2_{+} = 1.44 + 1.31\, n + 0.288\, n^2$
\\
\begin{tabular}[b]{||c|c|c|c|c|c||}
\hline
1.89 & 3.83 & 6.31 & 9.34 & 12.9 & 17.1\\
\hline
21.9 & 27.2 & 33.1 & 39.5 & 46.6 & 54.2\\
\hline
62.4 & 71.2 & 80.6 & 90.5 & & \\
\hline
\end{tabular} &
\begin{tabular}[b]{||c|c|c|c|c|c||}
\hline
3.01 & 5.20 & 7.96 & 11.3 & 15.2 & 19.7\\
\hline
24.7 & 30.3 & 36.5 & 43.3 & 50.7 & 58.6\\
\hline
67.1 & 76.2 & 85.9 & 96.1 & & \\
\hline
\end{tabular}
\end{tabular}
\end{center}
\end{table}

\begin{figure}[htb]
\begin{minipage}[b]{0.7\linewidth}
\begin{center}
\includegraphics[width=10cm]{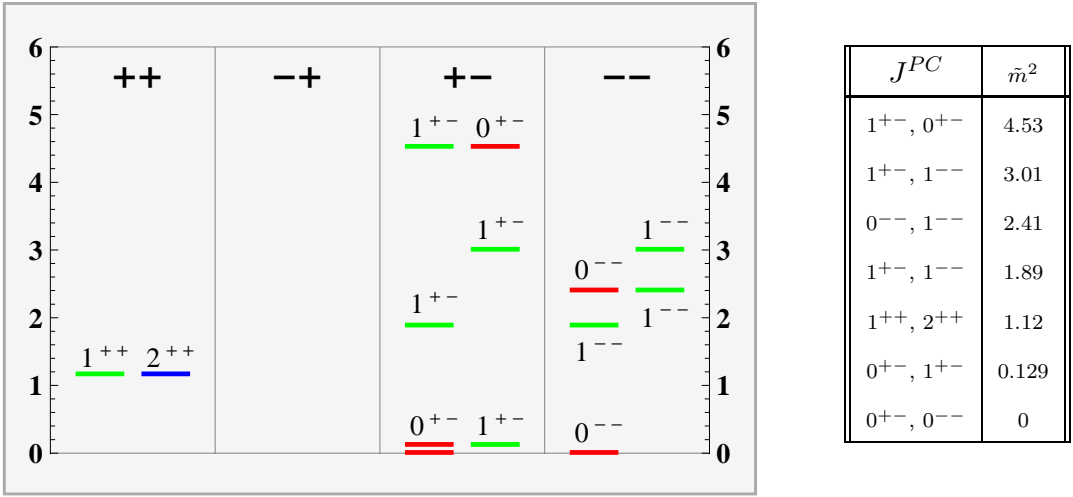}
\end{center}
\end{minipage}
\hspace{0.2cm}
\begin{minipage}[t]{0.2\linewidth}
\begin{center}
\vspace{-6cm}
{\scriptsize
\begin{tabular}{||c|c||}
\hline
\small $J^{PC}$ & $\mt^2$ \\
\hline
$1^{+-}$, $0^{+-}$ & 4.53 \\
$1^{+-}$, $1^{--}$ & 3.01  \\
$0^{--}$, $1^{--}$ & 2.41  \\
$1^{+-}$, $1^{--}$ & 1.89  \\
$1^{++}$, $2^{++}$ & 1.12  \\
$0^{+-}$, $1^{+-}$ & 0.129  \\
$0^{+-}$, $0^{--}$ & 0  \\
\hline
\end{tabular}
}
\end{center}
\end{minipage}
\caption{\small Values of $\tilde{m}^2$ and $J^{PC}$ quantum numbers of the states from the $\su(2)\times \su(2)$ invariant $\Ic$-odd sector. Each infinite tower is represented by it's lightest massive mode. Also in the figure: the massless scalar multiplet and the lightest states of the $\Ic$-even Graviton multiplet $1^{++},2^{++}$.}
\label{fig spectrum}
\end{figure}

In the figure \ref{fig spectrum} we collected the information about the spectrum of the $\Ic$-odd sector. It contains two massless scalars~\cite{GHK}, the lightest massive scalars from massive vector multiplets \cite{BDKS}, and lightest vectors from the seven vector towers discovered in this work. We have also added to the figure two $\Ic$-even bosonic states from the lightest graviton multiplet, a tensor $2^{++}$ state~\cite{DM1} and a vector $1^{++}$ dual to the $U(1)_{\cal{R}}$ current~\cite{DM}. These states share the spectrum of the ``minimal'' scalar and hence the lowest mass of their spectrum is a natural reference point. More $\Ic$-even scalar glueballs were found in the works~\cite{BHM1,BHM2}.

The quantum numbers of the $\Ic$-odd scalars from figure \ref{fig spectrum} were identified in \cite{BDKS}. The massless states are a scalar and a pseudoscalar; $0^{+-}$ and $0^{--}$. The corresponding tower of massive states is described by a vector multiplet, which contains a scalar $0^{+-}$ and a pseudovector $1^{+-}$. The latter mixes with another massive vector multiplet from the ansatz (\ref{ansVH2}), (\ref{ansVF2}) and (\ref{deltaF5_2}). Hence both of them should have the same quantum numbers from above $1^{+-}$. The vector state from the vector multiplet described by (\ref{ansVH1}), (\ref{ansVF1}) and (\ref{deltaF5_1}) have opposite parity transformations and therefore describes the $1^{--}$ vector state. One can draw the same conclusion by looking at the supermultiplet structure: this vector lies in the same supermultiplet with the pseudoscalar $0^{--}$.  The quantum numbers of the remaining four vectors are straightforward. The ones described by (\ref{ansVH2})-(\ref{deltaF5_2}) are pseudovectors $1^{+-}$ and the other two from (\ref{ansVH1})-(\ref{deltaF5_1}) are vectors $1^{--}$.

Two mixing vector multiplets consisting of the $0^{+-}$ scalar and the $1^{+-}$ vector correspond to the operators of different dimensions. Therefore their spectra are significantly different. To identify the spectra we associate the lighter modes with the operators of lower dimensions. Thus following~\cite{BDKS}, we identify the lightest massive multiplet in the figure~\ref{fig spectrum} to correspond to the $U(1)_{\rm Baryon}$ current (Betti) multiplet, which contains a scalar and a vector of dimensions 2 and 3 respectively.

As seen from the figure~\ref{fig spectrum}, the states from the Betti multiplet are much lighter than the other glueballs from the $\Ic$-odd sector and the known states from the $\Ic$-even sector. It would be interesting to compare the mass of the lightest state from the Betti Multiplet  with the mass of the lightest glueball created by the chiral operator ${\rm Tr}(AB)$. Despite a charge  under the $\su(2)\times \su(2)$ symmetry, the latter has the lowest dimension in the KS theory; $\Delta=3/2$. Therefore the corresponding state is a natural candidate to be the lightest massive mode in the KS spectrum.

\subsection{Scaling dimensions and SQM}
The KS solution explicitly breaks both conformal and $U(1)_{\cal R}$ symmetries. Therefore the fluctuations with different scaling dimensions and $\cal{R}$-charges can mix with each other. Indeed we saw earlier in section \ref{4-system sec} that the uncharged Betti vector mixes with the perturbation of the R-R four-form which carries $U(1)_{\cal R}$-charge $\pm 2$. Similarly the scalar of dimension 2 mixes with the scalar of dimension 5 in (\ref{ansHeq1})-(\ref{ansHeq2}).

The mixing between different multiplets of different dimensions can confuse the dimension analysis. Namely one cannot derive the dimension of the mode by merely analyzing the corresponding equations of motion in the large $\tau$ limit as it is usually done in the conformal case. A proper choice of basis fluctuations may be required to identify the corresponding multiplet structure and the dimensions. To illustrate this point we consider an example of the decoupled vector multiplet.

In section 3 the scalar particle $\tilde{\chi}$ described by (\ref{ansFeq}) was found to be degenerate with the vector fluctuation $\At$ that satisfies the same equation (\ref{Aeq_same}). Clearly both states must belong to the same $j=1/2$ multiplet. As they  satisfy the same equation the naive large $\tau$ analysis implies that they have the same dimension $\Delta=5$. This must be wrong as the bosonic states from the $j=1/2$ multiplet have the dimensions $\Delta_{1},\Delta_{0}$ that differ by $\Delta_1-\Delta_0=1$.

To resolve the puzzle we notice that the vector $\At$ mixes with other degrees of freedom, namely $\H$ and $\Nt$. In section 3 we chose $\At$ to be an independent variable, but we can choose $\H$ to be an independent variable instead ($\Nt$ cannot be chosen as an independent variable as it vanishes in this case). After eliminating $\At$ and redefining $\Ht=\sqrt{I} K\sinh\t\, \H$ the system (\ref{Nteq})-(\ref{Heq}) reduces to the equation
\bea
\label{sqme}
\Ht^{\prime\prime}+ \left(\frac{1}{2}\,\frac{I^{\prime\prime}}{I}-\frac{(K\sinh\tau)^{\prime\prime}}{K\sinh\tau}+\frac{I^\prime}{I}\frac{(K\sinh\tau)^\prime}{K\sinh\tau}-\frac34\,\frac{{I^\prime}^2}{I^2}\right)\Ht + \mt^2\, \frac{I}{K^2}\,\Ht =0\ .
\eea
At the large $\tau$ limit this equation behave as
\bea
\Ht'' - \frac{16}{9}\,\Ht\simeq 0\ ,
\eea
which indicates that $\H$ has dimension $\Delta=6$, in accordance with the $j=1/2$ multiplet structure. This is exactly what we expected since $\H$ corresponds to the fluctuation $a_{\mu\nu}$ from table 1. The later indeed has dimension six.

Let us note that one cannot favor (\ref{sqme}) over  (\ref{Aeq_same}) without knowledge of the supermultiplet structure. In fact both equations (\ref{Aeq_same}) and (\ref{sqme}) possess the same spectrum as they can be related to each other by the Supersymmetric Quantum Mechanics (SQM) transformation. More precisely this means that there are two first order differential operators $Q_+$ and $Q_-$, such that $Q_+Q_- \psi=m^2 \psi$ gives the equation  (\ref{Aeq_same}), while $Q_-Q_+ \psi=m^2 \psi$ leads to (\ref{sqme}).
The SQM transformation $\psi\rightarrow Q\psi$ which turns the solution of one equation into the solution of another changes the dimension of the corresponding mode. For the multiplets with half-integer $j$ the bosonic states should have different
dimensions $|\Delta_{+-}-\Delta_{-+}|=1$ and the SQM transformation is a five-dimensional truncation of the ten-dimensional supersymmetry transformation.  Among explicit examples there are the $j=1/2$ multiplet considered in this paper and the graviton multiplet studied in \cite{DM}. The latter contains two bosonic states of dimension 3 and 4, and the corresponding equations are also related by a SQM transformation.

Our logic also suggests that in addition to the equations (\ref{B+eqIKS3})-(\ref{CeqIKS3}) there should be a SQM-related system of equations governing the dynamics of the vectors $\B_+,\Ct$ with the same spectrum and with the large $\tau$ behavior that corresponds to the correct dimensions $3$ and $6$. It would be interesting to find this system explicitly by choosing $\D$ as an independent variable instead of $\C$.

The bosonic states from the multiplets with integer $j$ have the same dimensions
and hence should be described by the same equation. Thus each $j=1$ multiplet containing vector $\X$ and axial vector $\Y$ is described by a single equation governing both particles.

\subsection{Operators of the dual gauge theory}
\label{operators sec}

In  section~\ref{KW modes sec} we explained how the four-dimensional massive multiplets discussed above are embedded in the structure of the superconformal multiplets of the KW theory~\cite{Ce1}. Namely they exhaustively match the spectrum of the shortened $\su(2)\times \su(2)$ singlet multiplets of Vector type I and Gravitino types II and IV. Let us remind the reader of the operators that correspond to those superconformal multiplets.

The Betti multiplet, which is the ``massless'' type I Vector Multiplet (here quotes indicate that massless refer to the five-dimensional mass), corresponds to the operator
\bea
\label{U-operator}
{\cal{U}}= \tr Ae^V\bar{A}e^{-V} - \tr Be^V\bar{B}e^{-V}.
\eea
The lowest component of this operator $\tr \left(A\bar A - B\bar B\right)$ is dual to the scalar $U$ \cite{DKS} and has dimension $\Delta=2$.

The complex type IV Gravitino multiplet corresponds to the operator
\be
\bar{L}^{2k}_{\dot{\alpha}} = \tr e^V\bar{W}_{\dot{\alpha}}e^{-V}W^2(AB)^k\ ,
\ee
where $k$ labels representations of the R-symmetry group. The lowest (spin 1/2) component of this operator has dimension $\Delta = 3/2\,k+ 9/2$. The $\su(2)\times \su(2)$ invariant sector corresponds to $k=0$. In this case the dependence on the bi-fundamental fields $A$ and $B$ vanishes
\bea
\label{operator}
{\cal{O}}=\tr e^V\bar{W}_{\dot{\alpha}}e^{-V}W^2\ .
\eea
This is very interesting as this operator belongs to the pure gauge ${\cal{N}}=1$ SYM sector of the dual field theory. For $k=0$ the Gravitino multiplets of types II and IV are similar to each other. In particular, the type II multiplet corresponds to the complex conjugate of the operator $L^{20}_{\alpha}$ (\ref{operator}).

The five-dimensional superconformal multiplets split into the irreducible representations of the superalgebra in four dimensions. We saw that the Gravitino~II and Gravitino~IV multiplets split into four towers of massive supermultiplets, from which the lightest ones are presented in the figure~\ref{fig spectrum}. Down the throat they mix with the Betti multiplet and with each other. This means that the dual operators mix with each other at low energies. It would be interesting to understand how this mixing affects the masses of the corresponding glueballs from the field theory point of view.

\section{Discussion and final remarks}
\label{discussion}
In this paper we discussed the $\Ic$-odd $\su(2)\times \su(2)$ invariant bosonic excitations over the KS solution. At the massless level there are two spin 0 zero states: a Goldstone pseudoscalar that corresponds to the spontaneously broken $U(1)_{\rm Baryon}$ and a scalar related to the expectation value of the baryon operators. Together with fermions these states form a $j=0$ scalar supermultiplet. At the massive level the supersymmetry representation changes so that the pseudoscalar is eaten by the Betti pseudovector giving rise to a tower of $j=1/2$ vector supermultiplets. In the conformal case the $j=1/2$ multiplets are embedded into the ``massless'' Vector Multiplet of type~I~\cite{Ce1}.

There are two more towers of massive spin 0 modes (scalar and pseudoscalar) and six more massive spin 1 towers (3 vector and 3 axial vector). In the conformal case they belong to a combination of the shortened Gravitino Multiplets~II and~IV.

The two massive scalar excitations  mix with each other while the massive pseudoscalar excitation decouples.
Similarly the seven massive (pseudo)vectors split into two non-interacting subsystems of three vectors and four axial vectors. The system of three vectors contains the superpartner of the only massive pseudoscalar and two vectors $\X_{\pm}$.
The system of four axial vectors contains two superpartners of the two coupled massive scalars and the two axial vectors $\Y_{\pm}$. The states $\X_+,\Y_+$ and $\X_{-},\Y_-$ are degenerate in pairs and form two $j=1$ ``gravitino'' multiplets that consist of a vector, an axial vector and the spin 1/2 and 3/2 fermions.

The spin 0 massive modes from the $\Ic$-odd sector were found and studied in \cite{BDKS}. In particular the spectra of the corresponding $j=1/2$ supermultiplets were calculated numerically. In this work we identified the anticipated vector superpartners of the spin~0 states together with the remaining $\Ic$-odd vector states and computed numerically the spectra of the two $j=1$ multiplets. The results for the lightest states together with their $J^{PC}$ quantum numbers were presented in the figure~\ref{fig spectrum}.

An interesting task for the future would be to generalize our analysis to the $\Ic$-even sector and identify all $\su(2)\times \su(2)$ invariant bosonic modes of the KS theory. Some $\Ic$-even states are already known. Among them are the vector and the spin two states from the Graviton multiplet (the lightest modes are shown in figure~\ref{fig spectrum}). In fact these states are likely to be the only bosonic non-scalar states in the $\su(2)\times \su(2)$ invariant $\Ic$-even sector. Indeed there are no spin~1 $\Ic$-even excitations of $B_2$ and $C_2$ and the only possible spin~1 fluctuations of the metric were considered in~\cite{BHM1} and~\cite{DM1}. Some of the scalar states, namely a system of seven $0^{++}$ excitations were studied by M.~Berg et al. in~\cite{BHM1,BHM2}. They calculated the spectra of the particles but did not identify the corresponding operators. Besides an obvious task to find the corresponding pseudoscalar superpartners it would be interesting to match the resulting supermultiplets to the superconformal multiplets of~\cite{Ce1}.

Comparing our results with those for a pure gauge non-supersymmetric theory may give a sensible prediction for the masses of some of the lightest $\Ic$-even scalars. As we observed above, some of the fluctuations considered in this paper are dual to the operators that do not contain the bi-fundamental fields $A$ and $B$. In particular, the graviton multiplet, which contains $1^{++}$ and $2^{++}$ states, is dual to the ``supercurrent'' operator $V_{\alpha\dot{\alpha}}=\tr  W_{\alpha}e^V\bar{W}_{\dot{\alpha}}e^{-V}$~\cite{FZ}.  Also the states of the Gravitino Multiplets correspond to the components of the superfield ${\cal{O}}=\tr e^V\bar{W}_{\dot{\alpha}}e^{-V}W^2$ in the conformal case. In the KS theory however, the latter mix with the states from the Betti multiplet, dual to $A$ and $B$ dependent operators. Below we plot the lightest states from the pure gauge sector of the KS theory (figure 2.a) and compare them with those of the pure $\su(3)$ theory (figure 2.b). In figure 2.a we employ a qualitative approach, ignoring the mixing between the states from the pure gauge sector (i.e. $A$ and $B$ independent) and from the KK sector (with $A$ or $B$).

\begin{figure}[htb]
\begin{minipage}[b]{0.5\linewidth}
\begin{center}

\includegraphics[width=7.8cm]{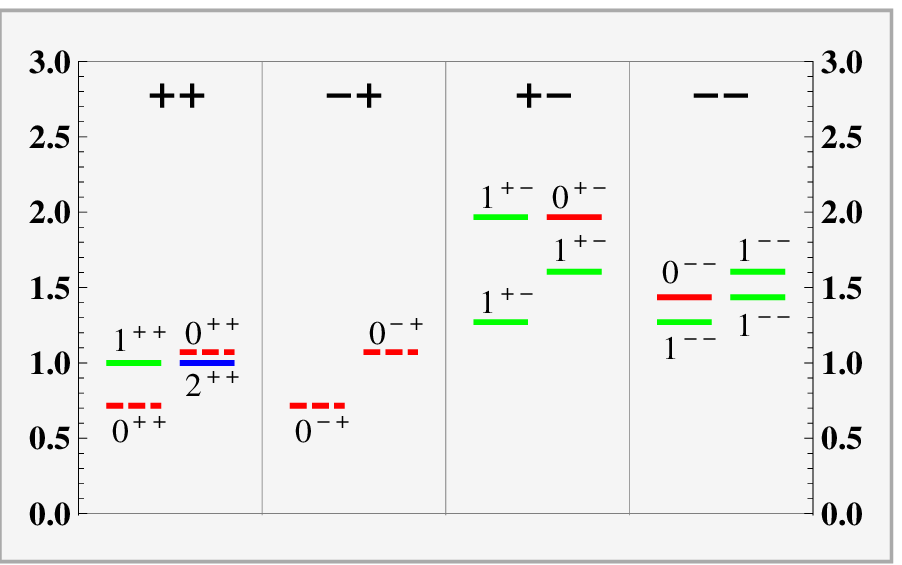}

(a)
\end{center}
\end{minipage}
\begin{minipage}[b]{0.5\linewidth}
\begin{center}

\includegraphics[width=7.8cm]{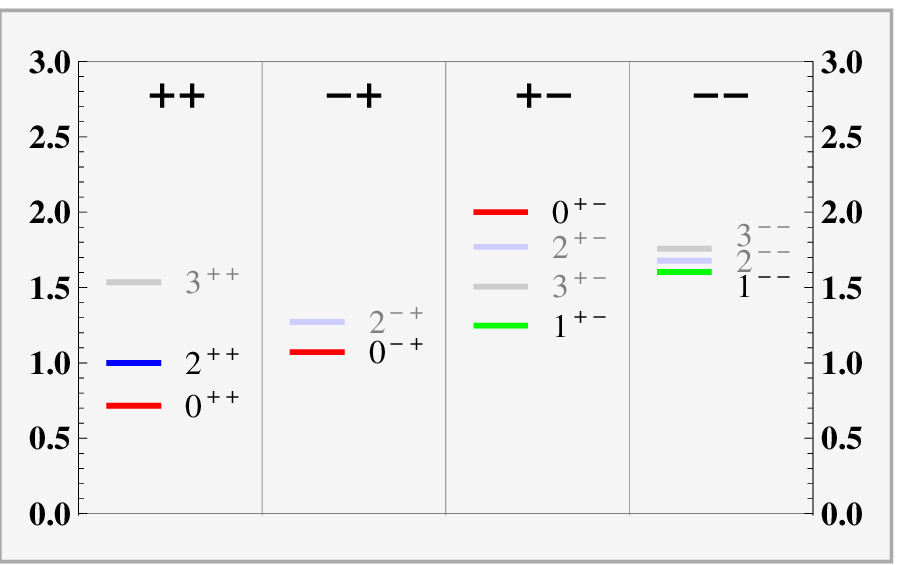}

(b)
\end{center}
\end{minipage}
\vspace{-0.6cm}
\caption{\small (a) Pure gauge sector of the KS theory. Conjectured positions of $0^{++}$ and $0^{-+}$ states are marked by dashed lines. (b) Spectrum of non-supersymmetric pure glue $\su(3)$ theory~\cite{Morningstar}. Both spectra are normalized to the mass of $2^{++}$ state.}
\label{fig SYM spectrum}
\end{figure}

In the figure~\ref{fig SYM spectrum}.a we present only those states from figure~\ref{fig spectrum}  that belong to the pure gauge sector of the KS theory. The masses of the states are normalized to the mass of the $2^{++}$ state. We have also plotted two light $\Ic$-even scalar multiplets, which we expect to see in the spectrum. These two multiplets should correspond to a mixture of the following pure $\Nn=1$ SYM operators: the gluino bilinear $\lambda\lambda$ of dimension $3$ and the dimension $4$ operators $\tr F_{\mu\nu}F^{\mu\nu}$ and $\tr F_{\mu\nu}\tilde{F}^{\mu\nu}$. These multiplets have not been identified yet and we mark their position with dashed lines.  Their masses in figure 2.a are conjectured based on the comparison with the pure glue $\su(3)$ theory.
It is also possible that some of the two $0^{++}$ particles in question is a part of the seven scalar system of~\cite{BHM1,BHM2}.

In the figure~\ref{fig SYM spectrum}.b we plot the lattice results of Morningstar and Peardon~\cite{Morningstar} for spectrum of the pure glue $\su(3)$ theory, which we also normalize to the mass of the $2^{++}$ state. We shade the irrelevant high spin states, which cannot be described in the supergravity approximation. Although the two theories are very different, the relative masses of the states are surprisingly similar. Indeed each state from the pure glue $\su(3)$ theory has a counterpart with the same quantum numbers and a similar mass (measured in the units of $2^{++}$ mass) in the pure gauge sector of the KS theory. Besides the counterparts of the pure glue $\su(3)$ theory states, the figure 2.a also contains their superpartners and even one ``extra'' vector multiplet (a $0^{--}$ scalar and a $1^{--}$ vector). In general the additional states are attributed to the fermionic degrees of freedom which are absent from the pure glue $\su(3)$ theory. Let us emphasize that the reason for the similarity between figure 2.a and 2.b is not immediately clear and could be coincidental. To examine this issue in more detail is an interesting problem for the future.

\vspace{0.5cm}

We are grateful to M.~Bianchi, M.~Douglas, A.~Hanany, Y.~Oz, M.~Shifman, M.~Strassler and especially I.~Klebanov for useful discussions. The research of A.D. is supported by the Stanford Institute for Theoretical Physics, the NSF under grant PHY-0244728, the DOE under contract DE-AC03-76SF00515 and in part by Grant RFBR 07-02-00878, Grant for Support of Scientific Schools NSh-3035.2008.2. A.D. would like to thank the workshop ``From Strings to Things'' at the University of Washington where part of this work was done. The research of D.M. was supported in part by the DOE grant DE-FG02-96ER40949, grant RFBR 07-02-01161, the Grant for Support of Scientific Schools NSh-3035.2008.2, the center of excellence supported by the Israel Science Foundation (grant number 1468/06), the grant DIP H52 of the German Israel Project Cooperation and the BSF United-States-Israel binational science foundation grant 2006157, and German Israel Foundation (GIF) grant No.~962-94.7/2007. The work of A.S. was supported by NSF grant No.~PHY-0756966.

\newpage
\appendix

\section{Useful facts about KS background}
\label{app:KS}
Here we present some useful information about the KS solution. We follow the notations of \cite{GHK,BDKS} and set $g_S= \alpha'=1$ and $M=2$.

We start with listing the external differentials for the $\su(2)\times \su(2)$ invariant forms on $T^{1,1}$
\bea
\d g^5 &=& - \bigl( g^1 \wg g^4 + g^3 \wg g^2 \bigr) \,,
\\
\d (g^1\wg g^3 + g^2 \wg g^4) &=& (g^1 \wg g^2 - g^3 \wg g^4) \wg g^5 \,.
\\
\d (g^1 \wg g^2) &=& - \half \bigl( g^1\wg g^3 + g^2\wg g^4 \bigr) \wg g^5 \,,
\\
\d (g^3 \wg g^4) &=& - \d (g^1 \wg g^2) \,.
\eea

The NSNS two-form of the KS solution and the corresponding field strength are
\bea
B_2 &=& f(\tau) g^1\wedge g^2
+  k(\tau) g^3\wedge g^4 \,,
\\
H_3 = \d B_2 &=& \d\tau\wedge (f' g^1\wedge g^2
+  k' g^3\wedge g^4) + \half\, (k-f)
g^5\wedge (g^1\wedge g^3 + g^2\wedge g^4) \,,
\eea
while the RR three-form field strength is
\bea
F_3 &=& g^5\wedge g^3\wedge g^4 + \d [ F(\tau)
(g^1\wedge g^3 + g^2\wedge g^4)]
\\  \nonumber
&=& g^5\wedge g^3\wedge g^4 (1- F)
+ g^5\wedge g^1\wedge g^2 F + F' \d\tau\wedge
(g^1\wedge g^3 + g^2\wedge g^4) \,.
\eea
They are defined with help of the auxiliary functions
\bea
F(\tau) &=& \frac{\sinh \tau -\tau}{2\sinh\tau}\,,
\nonumber \\
f(\tau) &=& \frac{\tau\coth\tau - 1}{2\sinh\tau}\,(\cosh\tau-1) \,,
\\ \nonumber
k(\tau) &=& \frac{\tau\coth\tau - 1}{ 2\sinh\tau}\,(\cosh\tau+1) \,,
\eea
which satisfy some useful identities like
\bea
k-f &=& 2\, F^\prime \,,
\\
f^\prime &=& (1-F)\, \tanh^2(\tau/2) \,,
\\
k^\prime &=& F\, \coth^2(\tau/2) \,.
\eea

Following \cite{KS} we also introduce the function $\ell(\t)$ via
\bea
F_5 &=& (1+\ast) B_2 \wg F_3 \;=\; (1+\ast)\, \ell(\t) \o_2 \wg \o_3 \,.
\eea
It is convenient to express it through the auxiliary functions from above
\bea
\ell (\t) &=& 2f + 4FF' \; \equiv \; 2f (1-F) + 2kF \,.
\eea

The metric of the deformed conifold is
\bea
\d s^2=\frac{\epsilon^{4/3}K}{2}\left(\frac{\d\tau^2+(g^5)^2}{3K^3}+\sinh^2\left(\frac{\tau}{ 2}\right)\left((g^1)^2+(g^2)^2\right)+\cosh^2\left(\frac{\tau}{2}\right)\left((g^3)^2+(g^4)^2\right)\right).
\eea
The inverse metric components written in the $\d\tau, g^1,..,g^5$ basis are
\bea
G^{11} = G^{22} &=& \frac{2}{\epsilon^{4/3} K(\tau)\sinh^2 (\tau/2) h^{1/2}(\tau)} \,,
\\
G^{33} = G^{44} &=& \frac{2}{\epsilon^{4/3} K(\tau)\cosh^2 (\tau/2) h^{1/2}(\tau)} \,,
\\
G^{55} = G^{\tau\tau} &=& \frac{6\, K(\tau)^2}{\epsilon^{4/3} h^{1/2}} \,,
\\
\sg &=& \frac{\eps^4}{96}\, \h{1/2} \sinh^2\t \,.
\eea
Here
\be
K(\tau)= \frac{ (\sinh (2\tau) - 2\tau)^{1/3}}{ 2^{1/3} \sinh \tau}
\,,
\ee
and the warp factor is
\bea
h(\tau) = 4 \cdot 2^{2/3} \eps^{-8/3} I(\t)\,,
\\
I(\tau) \equiv
\int_\tau^\infty \d x \,\frac{x\coth x-1}{\sinh^2 x}\, (\sinh (2x) - 2x)^{1/3} \,.
\eea
Hence
\bea
h'(\t) &=& - 16 \eps^{-8/3} F'(\t) K(\t) \,.
\eea

Some useful relations between the metric components include:
\bea
h \sg (\ug{11})^2 \ug{55} &=& \coth^2 \thalf \,,
\\
h \sg (\ug{33})^2 \ug{55} &=& \tanh^2 \thalf \,,
\\
h \sg \ug{11} \ug{33} \ug{55} &=& 1 \,,
\\
f' (\ug{11})^2 &=& (1-F) \ug{11} \ug{33} \,,
\\
k' (\ug{33})^2 &=& F \ug{11} \ug{33} \,;
\eea
\bea
\h{1/2} \sg (\ug{55})^2 &=& \frac{3\eps^{4/3}}{8}\, K^4 \sinh^2 \t \,,
\\
\h{1/2} \sg \ug{11} \ug{33} &=& \frac{\eps^{4/3}}{6K^2} \,,
\\
h \sg \ug{55} &=& \frac{\eps^{8/3}h}{16}\, K^2 \sinh^2 \t\,,
\\
h^{3/2} \sg &=& \frac{\eps^4 h^2}{96}\, \sinh^2 \t \,,
\\
\goverh &=& \frac{6K^2}{\eps^{4/3}h} \,.
\eea

\paragraph{$\Ic$-symmetry.}
$\Ic$-symmetry is the $\Zbb_2$-symmetry of the KS solution. It interchanges the two spheres $(\theta_1,\phi_1)$ and $(\theta_2,\phi_2)$  and changes the sign of $F_3$ and $H_3$. Its action on the $\su(2)\times \su(2)$ invariant forms  is as follows:
\bea
g^5 &\to& g^5 \,,
\\
\d g^5 &\to& \d g^5 \,,
\\
g^1 \wg g^2 &\to& - g^1 \wg g^2 \,,
\\
g^3 \wg g^4 &\to&- g^3 \wg g^4 \,,
\\
g^1 \wg g^3 + g^2 \wg g^4 &\to& - (g^1 \wg g^3 + g^2 \wg g^4) \,.
\eea

\section{Derivation of the Linearized Equations}
\label{app;derivation}
Let us first make a small digression about our conventions. We choose the names for the forms in the ansatz so as to possibly keep the similarity with notations used in the similar calculation for the KT limit in~\cite{BDKS}. The 1-forms (vectors) are shown in boldface. We work with the $(-+++)$ Minkowski signature. The four dimensional operations such as the Hodge star $\ast_4$ and Laplacian $\lpl$ are performed w.r.t.\ the standard Minkowski metric (without the warp factor). As it was explained, the four dimensional one-forms are all divergence free:
\bea
\d_4 \ast_4 \form{F} &=& 0 \,.
\eea
The eigenvalue of the 4-Laplacian $\lpl$ is $m_4^2$, however for compactness we shall express all our formulae in terms of the dimensionless combination $\mt^2$:
\bea
\label{mt}
m_4^2 &=& \frac{3\, \eps^{4/3}}{2\cdot 2^{2/3}}\, \mt^2 \,.
\eea

\subsection{3-Vector System}
\label{3-system}
With the ansatz~(\ref{ansVH1}), (\ref{ansVF1}) and~(\ref{deltaF5_1}), Bianchi identity for $F_5$ at the linear order in perturbation leads to four independent equations when written in components. Those are
\bea
\label{Bia1eq1}
\half \,\K - \half\, \L + \M' + \N &=& - F' (\A+\E) \,,
\\
\label{Bia1eq2}
h\, \sg \ug{55} \bigl( (\ug{11})^2 \K + (\ug{33})^2 \L \bigr) &=& \H \,,
\\
\label{Bia1eq3}
h\, \sg (\ug{33})^2 \ug{55} \lpl \L - \h{1/2} \sg \ug{11} \ug{33} (\ug{55})^2 \N &=& F \lpl \H \,,
\\
\label{Bia1eq4}
\left[ h \sg (\ug{33})^2 \ug{55} \L \right] ' - h \sg \ug{11} \ug{33} \ug{55} \M &=& F \H' \,.
\eea
Equations of motion for $F_3$ give the two equations:
\bea
\label{F1eq1}
- 2h \sg \ug{55} \lpl \E &=& 2 (k-f) \h{1/2} \sg \ug{11} \ug{33} (\ug{55})^2 \N + \ell\, \lpl \H \,,
\\
\label{F1eq2}
\nn
\left[ 2h\sg \ug{55} \E \right]' &=& - 2h \sg \ug{55} \bigl( f'(\ug{11})^2 \K + k'(\ug{33})^2 \L \bigr) -
\\ && \qquad\qquad - 2 (k-f) h \sg \ug{11} \ug{33} \ug{55} \M - \ell\, \H' \,.
\eea
Another pair of equations appear from $H_3$ equation of motion:
\begin{multline}
\label{H1eq1}
\left[ \h{1/2} \sg (\ug{55})^2 \A' \right]' - 2\h{1/2} \sg \ug{11} \ug{33} \A + h\sg \ug{55} \lpl \A =
\\ =  -2 F' \h{1/2} \sg \ug{11} \ug{33} (\ug{55})^2 \N \,,
\end{multline}
\bea
\label{H1eq2}
\left[ 2h\sg \ug{55} \H' \right]' + 2\h{3/2} \sg \lpl \H &=& 2(1-F)\K + 2F\L + 4F' \M - \ell\, \E \,.
\eea
No other supergravity equations contribute. In fact, some equations in the system~(\ref{Bia1eq1})-(\ref{H1eq2}) are algebraic and can be solved for the functions $\E$, $\K$, $\L$, $\M$ in terms of the functions $\N$ and $\H$. After doing so and redefining $\N$ according to~(\ref{Nt}), one can notice that equation~(\ref{F1eq2}) becomes an identity. Thus, there are only three independent second order differential equations for three unknown functions $\Nt$, $\H$ and $\A$. Introducing $\At = K^2 \sinh\t\, \A$, those reduce to the system (\ref{Nteq}), (\ref{Ateq}), (\ref{Heq}).

As mentioned in the section~\ref{3-system ansec} to separate the eigenmodes one can first impose $\Nt=0$. Then the remaining equations for $\H$ and $\At$ are equivalent. After setting $\Nt=0$, the equation (\ref{Nteq}) becomes the first order equation~(\ref{HAt}). Using it, one can eliminate the first and second derivatives of $\H$ from (\ref{Heq}) and express $\H$ in terms of $\At$ and its derivative. This reduces the system to just one equation~(\ref{Aeq_same}). Let us stress that in this case the ansatz for $\dlt F_5$ simplifies,
\bea
\dlt F_5 &=& (1+\ast)\, \d_4 \H \wg H_3 \,;
\eea
which gives a natural generalization of the KT limit ansatz in~\cite{BDKS} to the complete KS background (recall that in the KT limit $H_3\sim \d\t \wg \o_2$).

To extract the remaining two modes the equations~(\ref{Nteq})-(\ref{Heq}) can be written in the following form (we have done the trivial rescaling $\Ht\to 2^{7/6}\Ht$, $\At\to 2^{7/6}\mt\At$):
\bea
\label{Ateq1}
\At^{\prime\prime} - \frac{\cosh^2\tau+1}{\sinh^2\tau}\,\At + \mt^2\,\frac{I}{K^2}\,\At -\frac{2^{-1/6}\mt I^\prime}{K^3\sinh\tau}\,\Nt &=&0 ,\qquad
\\
\label{Hteq1}\nn
\Ht^{\prime\prime} + \left(\frac12\,\frac{I^{\prime\prime}}{I}-\frac{(K\sinh\tau)^{\prime\prime}}{K\sinh\tau}+ \frac{I^\prime}{I}\,\frac{(K\sinh\tau)^\prime}{K\sinh\tau} - \frac{3}{4}\,\frac{{I^\prime}^2}{I^2}\right)\Ht + \mt^2\,\frac{I}{K^2}\,\Ht + &&
\\ + \frac{2^{-1/6}\sqrt{I}}{K\sinh\tau}\,\left(\frac{I^\prime\Nt}{IK}\right)^\prime &=& 0,\qquad
\\
\label{Nteq1}\nn
\Nt^{\prime\prime} - \left(\frac{\cosh^2\tau+1}{\sinh^2\tau} + \frac{{2^{-1/3}I^\prime}^2}{IK^4\sinh^2\tau}\right)\Nt + \mt^2\,\frac{I}{K^2}\,\Nt - \frac{2^{-1/6}I^\prime}{IK}\,\left(\frac{\sqrt{I}\Ht}{K\sinh\tau}\right)^\prime - &&
\\ - \frac{2^{-1/6}\mt I^\prime}{K^3\sinh\tau}\,\At  &=& 0.\qquad \
\eea
It follows from above that the three vectors $\At$, $\Ht$, $\Nt$ are collinear. Therefore it suffices to consider the three scalar equations for the three variables $A$, $H$, $N$. The problem reduces to finding the spectrum of the Hamiltonian $\Hc$,
\bea
\label{Hamiltonian}\nn
- \Hc \left(\ba[c] A \\ H \\ N \ea \right) &=&
\left(\ba[c]
A^{\prime\prime} - \frac{\cosh^2\tau+1}{\sinh^2\tau}\,A -\frac{2^{-1/6}\mt I^\prime}{K^3\sinh\tau}\,N
\\
H^{\prime\prime} + \left(\frac12\,\frac{I^{\prime\prime}}{I}-\frac{(K\sinh\tau)^{\prime\prime}}{K\sinh\tau}+ \frac{I^\prime}{I}\,\frac{(K\sinh\tau)^\prime}{K\sinh\tau} - \frac{3}{4}\,\frac{{I^\prime}^2}{I^2}\right) H + \frac{2^{-1/6}\sqrt{I}}{K\sinh\tau}\,\left(\frac{I^\prime N}{IK}\right)^\prime
\\
N^{\prime\prime} - \left(\frac{\cosh^2\tau+1}{\sinh^2\tau} + \frac{{2^{-1/3}I^\prime}^2}{IK^4\sinh^2\tau}\right) N - \frac{2^{-1/6}I^\prime}{IK}\,\left(\frac{\sqrt{I} H}{K\sinh\tau}\right)^\prime - \frac{2^{-1/6}\mt I^\prime}{K^3\sinh\tau}\, A
\ea\right) \,.
\eea
Let us stress that this Hamiltonian is Hermitian w.r.t.\ the inner product
\bea
\langle 1|2 \rangle &=& \int_0^\infty \!\d\t\, \frac{I}{K^2} \bigl( A_1 A_2 + H_1 H_2 + N_1 N_2 \bigr) \,,
\eea
and the mass eigenvalues are found from the equation
\bea
\label{spectral_problem}
\Hc \left(\ba[c] A \\ H \\ N \ea \right) &=& \mt^2\, \frac{I}{K^2} \left(\ba[c] A \\ H \\ N \ea \right) \,.
\eea
As a consequence, different eigenvectors are orthogonal with the weight $I/K^2$.

We have found the decoupled mode which corresponds to setting $N\equiv 0$. This corresponds to the subspace of the form
(see equation (\ref{Nteq1})):
\bea
( A\,, H\,, N) &=& \left( - \frac{K^2\sinh\tau}{\mt I}\,\left(\frac{\sqrt{I} H}{K\sinh\tau}\right)^\prime \,, H\,, 0 \right) \,.
\eea
It is natural to suggest that the two remaining modes $(\Ah,\Hh,\Nh)$ belong to the orthogonal complement of this subspace. Namely,
\begin{multline}
\int\! \d\tau \,\frac{I}{K^2}\left(- \Ah\frac{K^2\sinh\tau}{\mt I}\,\left(\frac{\sqrt{I}H}{K\sinh\tau}\right)^\prime + \Hh\, H\right) =
\\
\int\! \d\tau \left( \frac{\sqrt{I}}{K\sinh\tau}\,\left(\frac{\Ah\sinh\tau}{\mt}\right)^\prime + \frac{I}{K^2}\, \Hh\right)H =0.
\end{multline}
The latter is satisfied by
\bea
\mt \Hh = - \frac{K}{\sqrt{I}\sinh\tau}\,\left(\Ah\sinh\tau\right)^\prime \,,
\eea
or
\bea
\label{constr1}
\Ah^\prime = -\mt\,\frac{\sqrt{I}}{K}\,\Hh - \coth\tau\, \Ah.
\eea
Using this expression one can eliminate all the derivatives of $A$ from (\ref{Ateq1}) and obtain another first order relation,
\bea
\label{constr2}
\Hh^\prime = -\left(\log \frac{\sqrt{I}}{K\sinh\tau}\right)' \Hh + \mt\,\frac{\sqrt{I}}{K}\,\Ah - \frac{2^{-1/6}I^\prime}{\sqrt{I}K^2\sinh\tau}\,\Nh.
\eea
Differentiating (\ref{constr2}) and eliminating $\Ah$ and $\Ah'$ using (\ref{constr1}) and (\ref{constr2}) one recovers the equation (\ref{Hteq1}) for $\Hh$. Thus the equation (\ref{Hteq1}) can be omitted from the system, and $\Hh$ can be expressed via $\Ah$ using (\ref{constr1}). After the elimination of $\Hh$ the system of the two equations (\ref{Ateq1}) and (\ref{Nteq1}) for $\Ah$ and $\Nh$
reproduces the system (\ref{Aheq1}), (\ref{Nheq1}).
As it is shown in the main text, these two equations decouple giving rise to the two modes $\X_\pm$.

\subsection{4-Vector System}
\label{4-system}

Similarly to the previous example the excitations (\ref{ansVH2}), (\ref{ansVF2}) and (\ref{deltaF5_2}) lead to the following linearized equations. The Bianchi identity gives five equations
\bea
\label{Bia2eq1}
-\half\, h \sg \ug{55} \Bigl( (\ug{11})^2 \P - (\ug{33})^2 \Q \Bigr) + \Bigl( h\sg \ug{11} \ug{33} \ug{55} \R \Bigr)' &=& F' \D' \,,
\\
\label{Bia2eq2}
- \Bigl( \h{1/2} \sg (\ug{33})^2 (\ug{55})^2 \G \Bigr)' + h\sg (\ug{33})^2 \ug{55} \lpl \Q &=& f' \lpl \D \,,
\\
\label{Bia2eq3}
- \Bigl( \h{1/2} \sg (\ug{11})^2 (\ug{55})^2 \F \Bigr)' + h\sg (\ug{11})^2 \ug{55} \lpl \P &=& k' \lpl \D \,,
\\
\label{Bia2eq4}
\F + \P' - \R &=& F \J + f' \C \,,
\\
\label{Bia2eq5}
\G + \Q' + \R &=& (1-F) \J + k' \C \,.\qquad\quad
\eea
A pair of equations come from the $F_3$ equation of motion:
\begin{multline}
\label{F2eq1}
\left[ \h{1/2} \sg (\ug{55})^2 \C' \right]' - 2\h{1/2} \sg \ug{11} \ug{33} \C +
\\ + h\sg \ug{55} \lpl \C = \h{1/2} \sg (\ug{55})^2 \bigl( f'(\ug{11})^2 \F + k'(\ug{33})^2 \G \bigr),
\end{multline}
\bea
\label{F2eq2}
\left[ 2h\sg \ug{55} \D' \right]' + 2\h{3/2} \sg \lpl \D &=& 2k' \P + 2f' \Q + 4F' \R + \ell \J \,;
\eea
and a pair of equations from the equation of motion for $H_3$:
\bea
\label{H2eq1}
2h \sg \ug{55} \lpl \J &=& 2\h{1/2} \sg (\ug{55})^2 \bigl( F(\ug{11})^2 \F + (1-F)(\ug{33})^2 \G \bigr) + \ell \lpl \D,\quad
\\
\label{H2eq2}\nn
\left[ 2h\sg \ug{55} \J \right]' &=& 2h\sg \ug{55} \bigl( F(\ug{11})^2 \P + (1-F)(\ug{33})^2 \Q \bigr) +
\\ && \qquad\qquad\qquad\qquad\qquad\quad + 4F' h \sg \ug{11} \ug{33} \ug{55} \R + \ell \D' \,.
\eea

As in the case of the previous ansatz, one of the equations is not independent and it is easy to demonstrate that any of the equations~(\ref{Bia2eq1})-(\ref{Bia2eq3}) or~(\ref{H2eq1})-(\ref{H2eq2}) can be eliminated. Thus, we obtain a system of eight equations for eight unknown forms. To write it in a more convenient form we introduce $\Ft$ and $\Gt$ as in~(\ref{Ft}) and~(\ref{Gt}).

We solve the algebraic equations for ansatz functions $\P$, $\Q$, $\R$ and $\J$, which we express in terms of the functions $\Ft$ and $\Gt$. The remaining four coupled second order differential equations are most conveniently written in terms of the functions $I$, $K$, $\sinh\tau$ and their derivatives. This way we obtain a system
\begin{multline}
\label{4-system1}
\Ft'' -  \left[ \frac{2}{\sinh^2\tau} + \half \right] \Ft + \mt^2 \frac{I}{K^2}\, \Ft + \half\, \Gt + \left(\frac12\,K^3\sinh\tau + 2^{-4/3}\frac{I^\prime}{K}\right)(\D' -  \J) =
\\ = \frac12\,K\Ct -\frac{2^{-4/3}I^\prime}{K^3 \sinh\tau}\, \Ct \,,
\end{multline}
\begin{multline}
\label{4-system2}
\Gt'' - \left[ \frac{2}{\sinh^2\tau} + \half \right] \Gt + \mt^2 \frac{I}{K^2}\, \Gt + \half\, \Ft + \left(\frac12\,K^3\sinh\tau - 2^{-4/3}\frac{I^\prime}{K}\right)(\D' - \J) =
\\ = \frac12\,K\Ct +\frac{2^{-4/3}I^\prime}{K^3 \sinh\tau}\, \Ct \,.
\end{multline}
\begin{multline}
\label{4-system3}
\Ct'' - \frac{\cosh^2\t+1}{\sinh^2\t}\, \Ct + \mt^2 \frac{I}{K^2} \,\Ct =2^{1/3}\,\mt^2K(\Ft+\Gt) -\mt^2\,\frac{I^\prime}{K^3 \sinh\tau}\,(\Ft-\Gt) \,,
\end{multline}
\begin{multline}
\label{4-system4}
\D'' + \Bigl( \log(IK^2\sinh^2\t) \Bigr)' \D' + \mt^2 \frac{I}{K^2}\, \D + \frac{(I^\prime K^2\sinh^2\tau)^\prime}{IK^2\sinh^2\tau}\,\D =
\\ = - \frac{I^\prime}{I}\,\J+ \frac{1}{IK^2\sinh^2\tau}\,  \left( 2^{1/3}K^3\sinh\tau(\Ft+\Gt) + \frac{I^\prime}{K}(\Ft-\Gt) \right)'  \,;
\end{multline}
where $\Ct = K^2 \sinh\t\, \C$, and $\mt$ is defined in (\ref{mt}). $\J$ is expressed in terms of given functions as follows:
\bea
\J&=& -\frac{I^\prime}{I}\,\D + \frac{2^{1/3}K}{I\sinh\tau}\,(\Ft+ \Gt) + \frac{I^\prime}{IK^3\sinh^2\tau}(\Ft-\Gt) \,.
\eea

The form of the equations in (\ref{4-system1})-(\ref{4-system4}) suggests that we introduce $\B_{\pm} = \Ft\pm\Gt$, so that the equations take the form (\ref{B+eq}), (\ref{B-eq}), (\ref{CeqIKS2}), (\ref{DeqIKS2}) and (\ref{Jeq2}).

The system of the equations (\ref{B+eq})-(\ref{DeqIKS2}) can be further reduced. We show that it can be split into the two decoupled pairs of equations by imposing the two different constraints, $\B_\pm=0$; each of them leading to a consistent reduction.

First, we set
\bea
\label{constraint2 ap}
\B_- &=& 0 \,;
\eea
then (\ref{B-eq}) implies
\bea
\label{Dprimeeq}
\D^\prime - \J = -\frac{1}{K^2\sinh\tau}\, \Ct.
\eea
Differentiating this equation, using (\ref{Jeq2}) and plugging it into the equation (\ref{DeqIKS2}), one gets, after eliminating $\D'$ via (\ref{Dprimeeq}), a simple relation
\bea
\label{Ctprimeeq ap}
\Ct^\prime = \mt^2 I\sinh\tau\,\D - \coth\tau\, \Ct.
\eea
Note that differentiating (\ref{Ctprimeeq ap}) and then eliminating the derivatives of $\Ct$ from (\ref{CeqIKS2}) we recover (\ref{Dprimeeq}) (and therefore (\ref{DeqIKS2}) as well).
Thus, the constraint (\ref{constraint2 ap}) singles out a consistent subsystem of the two equations:
\bea
\label{B+eqIKS3 ap}
\B_+'' -  \frac{2}{\sinh^2\tau}\,\B_+ + \mt^2 \frac{I}{K^2}\, \B_+  &=& 2K\Ct  \,,
\\
\label{CeqIKS3 ap}
\Ct'' - \frac{\cosh^2\t+1}{\sinh^2\t}\, \Ct + \mt^2 \frac{I}{K^2}\, \Ct &=&2^{1/3}\mt^2K\B_+  \,.
\eea
After a trivial rescaling of variables it reproduces the scalar equations (\ref{ansHeq1}) and (\ref{ansHeq2}).

To find the complementary pair of equations, one can instead set
\bea
\B_+ &=& 0 \,.
\eea
Equation (\ref{B+eq}) implies a first order constraint
\bea
\label{Dprimeeq2}
\D' &=& -\,\frac{I^\prime}{I}\,\D + \frac{I'}{I K^3\sinh^2\tau}\,\B_- + \frac{1}{K^2\sinh\tau}\, \Ct \,.
\eea
Using this equation one can eliminate the derivatives of $\D$ from (\ref{DeqIKS2}) and get the relation
\bea
\label{Ctprimeeq2 ap}
\Ct^\prime = - \mt^2 I\sinh\tau\,\D - \coth\tau\, \Ct.
\eea
Note that after eliminating the $\Ct$ derivatives from (\ref{CeqIKS2}) using this equation we recover (\ref{Dprimeeq2}) (and thus (\ref{B+eq}) and (\ref{DeqIKS2})). There remains a consistent subsystem of the two equations for $\B_-$ and $\Ct$, (\ref{B-eqIKS4}) and (\ref{CteqIKS4}). As it is shown in the main text, they can be further decoupled, yielding the two equations identical to (\ref{Xeqs}).

\end{document}